\documentclass[twocolumn]{bmcart}

\usepackage{hyperref,amsthm,amsmath,graphicx,gensymb,mathtools,geometry}
\usepackage[utf8]{inputenc} 
\usepackage[noadjust]{cite}

\usepackage[compact]{titlesec}  
\titlespacing{\section}{0pt}{20pt}{10pt}
\titlespacing{\subsection}{0pt}{20pt}{10pt}

\definecolor{linkColor}{rgb}{1,0,0}
\hypersetup{pdfborder={0 0 0},colorlinks=true,urlcolor=linkColor,citecolor=linkColor,allcolors=linkColor}

\startlocaldefs
\newcommand{ \ii }{\emph{i} \,}
\newcommand{\tensor}[1]{\mathbf{#1}}
\newcommand{\ft}{{\mathcal{F}}}

\endlocaldefs

\begin{document}
\begin{frontmatter}

\begin{fmbox}
\dochead{Research}


\title{A Fast Image Simulation Algorithm for \\ Scanning Transmission Electron Microscopy }

\author[
   addressref={aff1},
   corref={aff1},  
   email={cophus@gmail.com}   
]{\inits{CO}\fnm{Colin} \snm{Ophus}}


\address[id=aff1]{                  
  \orgname{National Center for Electron Microscopy, Molecular Foundry, Lawrence Berkeley National Laboratory}, 
  \street{1 Cyclotron Road}, 
  \city{Berkeley},                              
  \cny{USA}                                    
}
\end{fmbox}
{ \small
\href{mailto:cophus@gmail.com}{cophus@gmail.com} \\
National Center for Electron Microscopy, Molecular Foundry \\
Lawrence Berkeley National Laboratory, Berkeley, CA, USA
}

\begin{abstractbox}
\begin{abstract} 

Image simulation for scanning transmission electron microscopy at atomic resolution for samples with realistic dimensions can require very large computation times using  existing simulation algorithms. We present a new algorithm named PRISM that combines features of the two most commonly used algorithms, the Bloch wave and multislice methods. PRISM uses a Fourier interpolation factor $f$ that has typical values of 4-20 for atomic resolution simulations. We show that in many cases PRISM can provide a speedup that scales with $f^4$ compared to multislice simulations, with a negligible loss of accuracy. We demonstrate the usefulness of this method with large-scale scanning transmission electron microscopy image simulations of a crystalline nanoparticle on an amorphous carbon substrate.



\end{abstract}

\begin{keyword}
\kwd{Scanning Transmission Electron Microscopy}
\kwd{Electron Scattering}
\kwd{Image Simulation}
\end{keyword}

\end{abstractbox}
\end{frontmatter}


\section*{Introduction}

Transmission electron microscopy (TEM) is one of the most versatile and powerful experimental tools for imaging and diffraction of micrometer to sub-nanometer structures. The recent widespread adoption of hardware aberration correction has in particular enabled routine atomic resolution imaging of structures \cite{batson2002sub, rose2005prospects, dahmen2009background}. A more recent technical advance for TEM experiments is the use of direct electron detectors. These cameras have a much higher quantum efficiency than standard charge-coupled devices with a scintillator, and can also operate at much higher speeds \cite{mcmullan2014comparison, gautam2015analysis, park2015, tate2016high}. Direct electron detectors  have already created dramatic improvements in plane-wave TEM imaging experiments, especially single-particle biological cryo-EM studies \cite{li2013electron, nogales2016development, glaeser2016good}. These detectors have also enabled many new kinds of experiments for scanning transmission electron microscopy (STEM), where the electron probe is converged to very small dimensions and scanned across the surface of a sample, because the camera speed is high enough to record a full image of the diffracted probe at each probe position \cite{ophus2014recording}. Examples include nanobeam electron diffraction strain measurements \cite{ozdol2015strain, pekin2016optimizing}, orientation mapping of semi-crystalline polymers \cite{panova2016orientation}, and phase contrast imaging modes such as differential phase contrast \cite{shibata2012differential,tate2016high}, phase plate interferometry \cite{ophus2016efficient}, and ptychography \cite{yang2016simultaneous}. Each of these experiments can benefit by accompanying them with STEM simulations to aid in interpretation or validation of the results.




However, while the experimental capabilities of TEM and STEM have expanded, simulation methods have remained largely unchanged for some time. The two primary methods currently used for atomic-resolution simulations are Bloch wave calculations and the multislice method \cite{allen2003lattice,findlay2003lattice,kirkland2010advanced}. In the Bloch wave method, the electron wavefunction is defined using a basis set that satisfies the Schr\"{o}dinger equation inside the sample. For a perfect crystal, Bloch waves are stationary solutions with the same periodicity and symmetry as the crystalline lattice. After calculating the eigenvectors and eigenvalues of this basis set, the wavefunction at the entrance surface of the sample can be matched to the known electron probe coefficients, and then the resulting electron wavefunction can be computed everywhere (including the exit surface of the sample) \cite{bethe1928theorie, zuo2013electron}. This scattering calculation can be written compactly in a scattering matrix (often called the $\tensor{S}$-matrix) formalism \cite{kirkland2010advanced}. Bloch wave calculations are almost never used for imaging or diffraction simulations of large samples (beyond the several `unit cell' scale for crystalline materials) for two reasons; the first is that eigendecomposition of a non-sparse Bloch wave matrix large enough to accurately simulate image sizes $\geq1000^2$ pixels would take an impractically long time to compute.  The second is that the storage requirements of this scale of $\tensor{S}$-matrix is greater than a terabyte, and using it would require trillions of multiplication operations \cite{kirkland2010advanced}.

A more efficient formulation for large electron scattering simulations is the multislice algorithm \cite{cowley1957scattering}. In this method, the atoms of the simulated sample are divided up into infinitely thin slices along the beam direction. The resulting electron scattering is calculated by alternating between a transmission operator through each slice, followed by Fresnel propagation of the electron wave to the next slice. These operations can be performed efficiently in realspace and reciprocal space respectively, and so an efficient implementation of this method requires a forward and inverse Fourier transform at each step \cite{kirkland2010advanced}. The multislice algorithm is very efficient for plane-wave, conventional TEM image or diffraction simulations. It is much less efficient for STEM simulations consisting of thousands or millions of probe positions. This is because while the atomic scattering potential can be reused for all probe positions, the transmission and propagation steps must be repeated for each additional probe position. The scattering potential calculations can be performed very efficiently using look-up tables \cite{shukla2015unravelling, ophus2016efficient} or a point scattering method \cite{vandenbroek2015fdes}, but the slow part of the calculation is usually repeated for all probe positions \cite{kirkland2010advanced}. Many STEM studies such as high precision 2D measurements \cite{yankovich2014picometre,yu2016integrated,kim2016direct}, 3D atomic electron tomography \cite{xu2015three,yang2016deciphering}, and others \cite{johnson2017three}, make use of image simulations of many thousands of STEM probe positions. This requires long computation times, even with modern implementations of the multislice method \cite{vandenbroek2015fdes, barthel2012time, grillo2013stem_cell, allen2015modelling,  hosokawa2015benchmark, lobato2016progress, kirkland2016computation}. It is therefore desirable to develop an electron scattering simulation algorithm that shares the calculation burden between STEM probe positions in a more efficient manner than multislice simulation. Chen et al.\ have proposed one such method \cite{chen1995modification}, but it has not found widespread application. For a detailed discussion of the relationship between the Bloch wave and multislice simulation methods, we refer readers to the derivations of Allen, Findlay et al~\cite{allen2003lattice,findlay2003lattice}.

In this manuscript, we derive a more efficient algorithm for STEM simulations by combining aspects of the multislice and Bloch wave methods. We use the multislice method to directly calculate a subset of the rows of the $\tensor{S}$-matrix (corresponding to plane waves of various orientations), which is then used in a similar manner as Bloch wave calculations \cite{chen1995modification} to relate the output wavefunction to a given input. The key insight is that because highly-converged STEM probes decay to zero quickly with distance from the probe center position, they can be cropped out of the full $\tensor{S}$-matrix in a highly-accurate Fourier interpolation scheme. The algorithm presented here is referred to as the plane-wave reciprocal-space interpolated scattering matrix (PRISM) algorithm. We also compare the accuracy and computation time of the PRISM and multislice algorithms, and suggest some useful extensions of the PRISM method.


\section*{Theory and Methods}

\subsection*{The Multislice and Bloch Wave Methods}

For previously published TEM simulation methods, we will  briefly outline the required steps here. We refer readers to Kirkland for more information on these methods \cite{kirkland2010advanced}. We will also only describe the scattering of the electron beam while passing through a sample; probe-forming optics and the microscope transfer function mathematics are described in many other works. All elastic scattering TEM simulations aim to describe how an electron wavefunction $\psi(\vec{r})$ evolves over the 3D coordinates $\vec{r} = (x,y,z)$.  The evolution of the slow-moving portion of the wavefunction along the optical axis $z$ can be described by the Schr\"{o}dinger equation for fast electrons \cite{kirkland2010advanced}
\begin{equation}
    \frac{\partial \psi(\vec{r})}{\partial z} = 
    \frac{\ii \lambda}{4 \pi} {\nabla_{xy}}^2 \psi(\vec{r})
    + \ii \sigma V(\vec{r}) \psi(\vec{r}),
    \label{EquationSchrodinger}
\end{equation}
where $\lambda$ is the relativistic electron wavelength, ${\nabla_{xy}}^2$ is the 2D Laplacian operator, $\sigma$ is the relativistic beam-sample interaction constant and $V(\vec{r})$ is the electrostatic potential of the sample.

The Bloch wave method uses a basis set that satisfies Eq.~\ref{EquationSchrodinger} everywhere inside the sample boundary, which is assumed to be periodic in all directions. This basis set is calculated by calculating the eigendecomposition of a set of linear equations that approximate Eq.\ref{EquationSchrodinger} up to some maximum scattering vector $|q_{\rm{max}}|$. Then, for each required initial condition such as different STEM probe positions on the sample surface, we compute the weighting coefficients for each element of the Bloch wave basis set. Finally, the exit wave after interaction of the sample is calculated by multiplying these coefficients by the basis set. This procedure can be written in terms of a scattering matrix $\tensor{S}$ as \cite{kirkland2010advanced}
\begin{equation}
    \psi_f(\vec{r}) = \tensor{S} \; \psi_0(\vec{r}),
    \label{EquationBloch}
\end{equation}
where $\psi_0(\vec{r})$ and $\psi_f(\vec{r})$ are the incident and exit wavefunctions respectively. The Bloch wave method can be extremely efficient for very small simulations, where the field of view is on the scale of crystalline unit cells. High symmetry is also an asset for Bloch wave simulations, as we can limit the beam of plane waves (beams) included in the basis set to a small number. However, for a large STEM simulation consisting of thousands or even millions of atoms in the simulation, the $\tensor{S}$-matrix may contain billions or more entries, which requires an impractical amount of time to calculate the eigendecomposition. And, actually using Eq.~\ref{EquationBloch} many times for various electron probes could take a very long time.  Thus Bloch wave methods are typically only used for very small size STEM simulations.

The most commonly employed method for large STEM simulations is the multislice algorithm. The multislice method alternates between solving the two terms on the right hand side of Eq.~\ref{EquationSchrodinger}, for thin slices of thickness $t$ taken from the sample. The left term is interpreted as a Fresnel propagation operator, which can be efficiently applied in Fourier space as \cite{kirkland2010advanced}
\begin{equation}
    \Psi_{p+1}(\vec{q}) =  \Psi_p(\vec{q})
    \exp(- \ii \pi \lambda |\vec{q} \, |^2 t)
\end{equation}
where $\Psi(q) = \ft \{ \psi(r) \}$ is the Fourier transform of $\psi(\vec{r})$, $\vec{q}=(q_x,q_y)$ is the 2D coordinate vector for Fourier space, and the subscript $p$ refers to the slice index. The second  operator of Eq.~\ref{EquationSchrodinger} can be efficiently applied in real space as
\begin{equation}
    \psi_{p+1}(\vec{r}) =  \psi_p(\vec{r})
    \exp \left[ 
    \ii  \sigma V_p^{\rm{2D}}(\vec{r}) 
    \right],
\end{equation}
where $V_p^{\rm{2D}}(\vec{r})$ is the 2D electrostatic potential of all atoms inside slice $p$, integrated over the slice along the beam direction from the 3D potential. In practice, the atomic potentials are integrated into 2D potentials before the simulation, and then added directly to the slice potential, or applied using convolution \cite{vandenbroek2015fdes}. These two steps describe how the electron wavefunction evolves slice-by-slice until it has interacted with the entire sample, applied sequentially as
\begin{equation}
    \psi_{p+1}(\vec{r}) = 
    \ft^{-1} \left\{
    \ft \left\{
    \psi_p(\vec{r})
    e^{ 
    \ii  \sigma V_p^{\rm{2D}}(\vec{r}) 
    } \right\}
    e^{- \ii \pi \lambda |\vec{q} \,|^2 t}
    \right\},
    \label{EquationMultislice}
\end{equation}
where $\ft^{-1} \left\{ \right\}$ is the inverse Fourier transform. The Multislice method is simple to implement and very accurate, but is not very efficient for large scale STEM simulation. The reason is that although the atomic potentials can be re-used for different probe positions, the remainder of the calculation (using Eq.~\ref{EquationMultislice} to propagate each probe though the sample) must be run independently. While this problem is amenable to parallelization, none of the calculations are shared between different probe positions, or different probe parameters such as defocus, convergence angle or probe tilt. In the next section, we will show how a STEM simulation can be reformulated into an $\tensor{S}$-matrix approach, where the computational load of applying Eq.~\ref{EquationMultislice} can be shared between different probe configurations.


\subsection*{The PRISM Algorithm for STEM Simulations}

\begin{figure*}[htbp]
    \centering
        \includegraphics[width=6.0in]{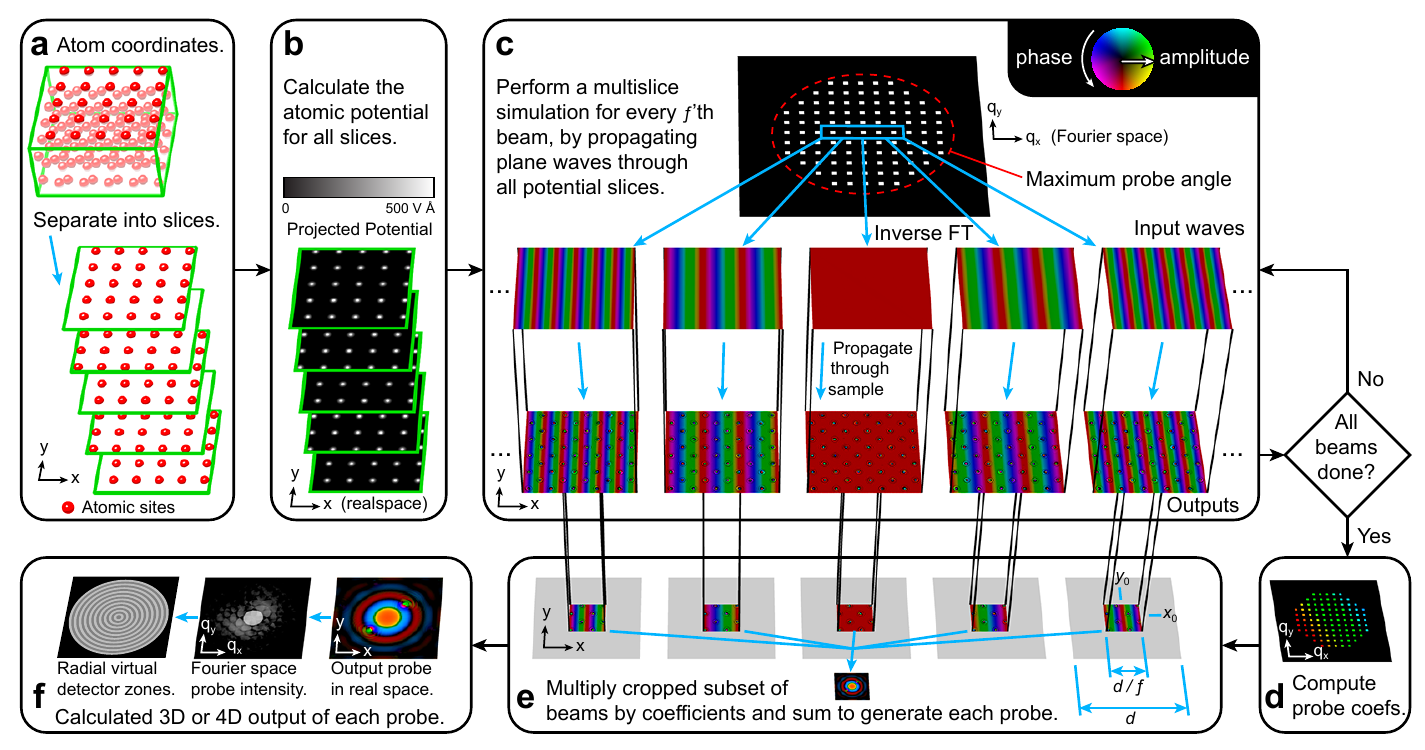}
 	\caption{The PRISM algorithm. (a) The sample's atomic coordinates are divided up into slices. (b) The projected potential of each slice is computed. (c) Each required plane wave is calculated by using the multislice algorithm to propagate the wave through the sample. (d) The complex coefficients for each probe are calculated for the interpolated / cropped coordinate system, and then (e) each wave is multiplied by the associated coefficient and summed to generate the probe. (f) Final probe wave is typically Fourier transformed and is either output as a CBED intensity or virtual detectors are used to add up subsets of the wave intensity.}
	\label{FigurePRISMflowchart}
\end{figure*}


The first step of the method proposed here is to separate all atomic coordinates of the simulation cell (which is assumed to be orthorhombic here) into slices, shown in Fig.~\ref{FigurePRISMflowchart}a. These slices can have unequal thickness to better match the atomic coordinates, but should not have thicknesses larger than the average atomic spacing as this could cause errors \cite{kirkland2010advanced}. The second step is to calculate the 2D projected potentials $V(\vec{r})$ for all slices, as in Fig.~\ref{FigurePRISMflowchart}b.

Next, we choose an interpolation factor $f$. In practice a different factor can be used in $x$ and $y$, but for simplicity we will describe the simulation method for a square (in the $(x,y)$ plane) simulation cell of size $d$. This factor $f$ should be chosen to be large enough so that a square area with a side length of the simulation cell size divided by $f$ can encompass all possible STEM probes after they pass through the cell.  This can be estimated by numerically simulating a few probes using the conventional multislice method or the method described here. We then also choose a maximum incident probe semi-angle $\alpha_{\rm{max}}$.  Note that the simulation will include larger scattering angles than this value, and that this value should be equal to the largest desired probe semiangle plus $f$ times the Fourier space pixel size $\Delta q$. We then determine a set of plane wave initial conditions to simulate using the multislice method, shown in Fig.~\ref{FigurePRISMflowchart}c.  This set of plane waves corresponds to the incident electron probe
\begin{eqnarray}
    \Psi_{m,n}(\vec{q}) = \delta (
    q_x - m f \Delta q, \;
    q_y - n f \Delta q
    ),
\end{eqnarray}
where $\sqrt{m^2+n^2} f \lambda 
\Delta q \leq \alpha_{\rm{max}}$, $\delta (\vec{q})$ is the delta function, and $(m,n)$ are integers representing the plane wave index. Thus, we compute only a subset of all possible periodic plane waves for the simulation cell size, reducing the number of waves calculated by a factor of $f^2$. These plane waves are stored in realspace in a large array that we will refer to as the compact $\tensor{S}$-matrix, with the output plane waves defined as $\tensor{S}_{m,n}(\vec{r})$. These output wave dimensions can be reduced by a factor of 4 if the multislice simulation used an anti-aliasing aperture position at half of the maximum scattering angle is used for the multislice simulations \cite{kirkland2010advanced}.

 
Next, we calculate each converged electron probe at position $\vec{r}_0 = (x_0,y_0)$ by first computing the required coefficients $\alpha_{m,n}(\vec{r}_0)$ for each plane wave $\tensor{S}_{m,n}(\vec{r})$, and then multiplying these coefficients by the associated plane wave basis and summing over a square sub-region with side length $d$ centered around the probe. This is shown schematically in Fig.~\ref{FigurePRISMflowchart}d. The sub-region is bounded by
\begin{eqnarray}
    && x_0 - \frac{d}{2 f} 
    \leq x < 
    x_0 + \frac{d}{2 f} \nonumber \\
    && y_0 - \frac{d}{2 f} 
    \leq y <
    y_0 + \frac{d}{2 f},
    \label{EquationRegionBounds}
\end{eqnarray}
giving a cutout region having an area of $d^2/f^2$, which should be periodically wrapped around the simulation cell boundaries. The wave coefficients are defined as
\begin{eqnarray}
    \alpha_{m,n}(\vec{r}_0) &=&
    A(\vec{q})
    \exp \left[-\ii  \chi(\vec{q}) \right]
    \nonumber \\
    && \exp 
    \left\{ -2 \ii \pi 
    \vec{q} \boldsymbol{\cdot} 
    \left[ \right. \right.
    x_0 - h \tan(\theta_x),
    \nonumber \\
    &&  y_0- h \tan(\theta_y)
    \left. \left. \right] \right\},
\end{eqnarray}
where $A(\vec{q})$ is the probe aperture function defined as
\begin{equation}
    \begin{array}{llll}
        A(\vec{q}) = & 1
        & 
        \rm{where} 
        & |\vec{q}| \leq q_{\rm{probe}} 
        \nonumber \\
        & 0 & \rm{elsewhere}. &
    \end{array}
\end{equation}

The probe can also contain coherent wave aberrations such as defocus $C_1$ or $3^{\rm{rd}}$ order spherical aberration $C_3$ described by the phase shift function \cite{kirkland2010advanced}
\begin{equation}
    \chi(\vec{q}) =
    \pi \lambda |\vec{q} \, |^2 C_1 
    + \frac{\pi}{2} \lambda^3 |\vec{q} \,|^4 C_3 
    + ...
\end{equation}
Finally, the terms $h \tan(\theta_x)$ and $h \tan(\theta_y)$ shift the probe back to the center of a cutout region for a given simulation cell of height $h$ and probe tilt angles $\theta_x$ and $\theta_y$. As shown in Fig.~\ref{FigurePRISMflowchart}e, once the probe coefficients $\alpha_{m,n}(\vec{r}_0)$ have been computed, the complex probe in realspace $\psi(\vec{r},\vec{r}_0)$ can be computed using the summation
\begin{equation}
    \psi(\vec{r},\vec{r}_0) =
    \sum_{m,n} \tensor{S}_{m,n}(\vec{r}) 
    \; \alpha_{m,n}(\vec{r}_0),
    \label{EquationPRISMfinal}
\end{equation}
in the cut out region defined by Eq.~\ref{EquationRegionBounds}. Note that this expression is simply an expanded form of Eq.~\ref{EquationBloch}. Eq.~\ref{EquationPRISMfinal} can be evaluated more quickly if we skip the addition of all terms where $\alpha_{m,n}(\vec{r}_0)=0$. After the probe is computed we can either output the full probe diffraction pattern, or more commonly integrate a subset of the probe intensity after taking its Fourier transform, as in Fig.~\ref{FigurePRISMflowchart}f. Once the output signals of all probes have been tabulated, the simulation is complete. Our method is very similar to that proposed by Chen et al.~\cite{chen1995modification}; But, where they include tilts of the various beams in the propagation operator, we have included it in the initial conditions of each beam, which negates the need for an offset term to relate the relative phases of the beams.

\subsection*{Simulation and Analysis Implementation}

\begin{figure*}[htbp!]
    \centering
        \includegraphics[width=6in]{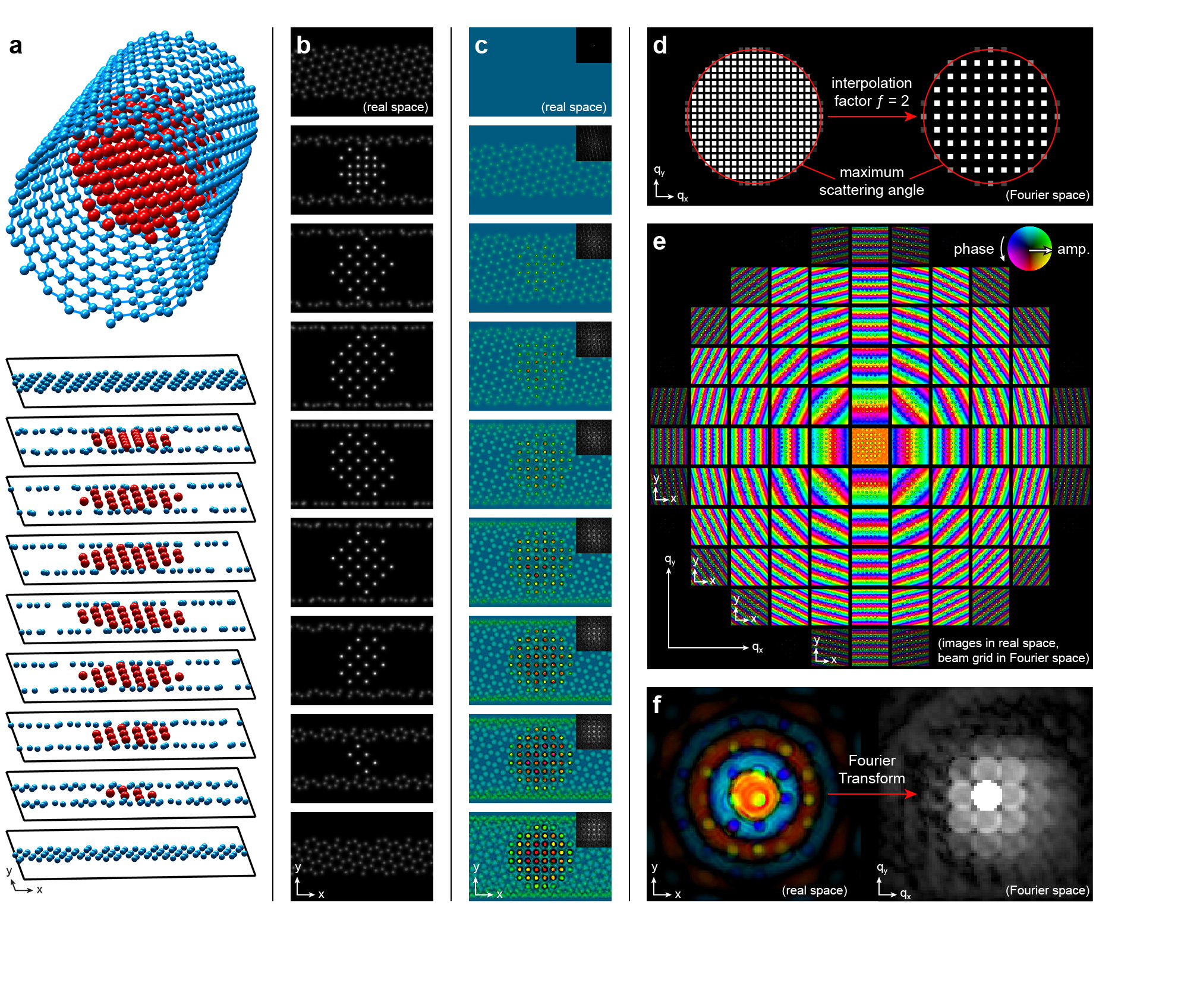}
 	\caption{Example implementation of the PRISM algorithm. (a) The sample's atomic coordinates are divided up into slices. (b) The projected potential of each slide is computed. (c) Each required plane wave is calculated by using the multislice algorithm to propagate the wave through the sample. (d) The complex coefficients for each probe are calculated for the interpolated / cropped coordinate system, and then (e) each wave is multiplied by the associated coefficient and summed to generate the probe. (f) Final output is typically the intensity of the probe's Fourier transform.}
	\label{FigurePRISMimplement}
\end{figure*}

All simulations and analysis in this study were performed using custom Matlab code. The multislice methods and the atomic potentials employed were taken from Kirkland \cite{kirkland2010advanced}.  Thermal scattering effects were implemented using the frozen phonon approximation, which involves repeating the calculation with different phonon configurations (approximated with random atomic displacements) and summing the results incoherently.

An implementation of the PRISM algorithm for a sample consisting of a nanoparticle contained within a carbon nanotube is shown in Figs.~\ref{FigurePRISMimplement}a-f. Each of the panels in this figure correspond to the same step as those given in Figs.~\ref{FigurePRISMflowchart}a-f. In Figs.~\ref{FigurePRISMimplement}c, e and f, the wave phase is shown as the color hue, while the wave amplitude is shown by the brightness of each pixel. All simulations were performed using a 80 kV accelerating voltage, a slice thickness of 0.2 nm, a pixel size of 0.01 nm, and we used no spherical aberration in the electron probes.

\subsection*{Calculation Time for PRISM Simulations}

We will now approximate the computation time of the PRISM algorithm, relative to traditional multislice simulations. We will neglect the computation time of the sample projected potential slices, as this calculation time is equal for both methods. We will also not consider thermal scattering, since it will require an increase in calculation time by an equal multiplier for both methods. For simplicity we will assume a square simulation cell with side length $N$ where $N$ is a power of two. Each slice will require the transmission and propagation operations given in Eq.~\ref{EquationMultislice}, which requires $6 N \log_2(N)$ complex operations for the forward and inverse Fourier transforms and $2 N^2$ operations to multiply the sample potential and the Fresnel propagation functions. If the entire STEM simulation consists of $P$ unique probe positions and $H$ slices through the sample, the total calculation time $T_{\rm{multi}}$ required is
\begin{eqnarray}
    T_{\rm{multi}} &=&
        H P \left[
        6 N \log_2(N) + 2 N^2
        \right] \nonumber \\
    &\approx& 2 H P N^2,
    \label{EquationTimeMulti}
\end{eqnarray}
if the simulation cell is large, i.e.~$N \gg 1$. The PRISM method requires two parts to compute the scattering of all STEM probes. The first half of the algorithm requires $B/f^2$ multislice simulations, where $B$ is the number of beams included in the full resolution simulation, which will be reduced by the interpolation factor squared. The second half is multiplication of the compact scattering matrix $S$ for all beams (multislice plane waves computed in the previous step), which is required for $P$ total probes, as in Eq.~\ref{EquationPRISMfinal}. This multiplication step is only required for the reduced number of beams $B/f^2$, and the cut out region defined by Eq.~\ref{EquationRegionBounds} will reduce the number of multiplication operations to $N^2 / 4 f^2$ (note the extra factor of $1/4$ is due to storing only the part of $S$ inside the anti-aliasing aperture). Therefore the total calculation time $T_{\rm{PRISM}}$ required for PRISM is
\begin{eqnarray}
    T_{\rm{PRISM}} &=&
        \frac{H B}{f^2} \left[
        6 N \log_2(N) + 2 N^2
        \right] 
        + \frac{P B N^2}{4 f^4}
        \nonumber \\
    &\approx& B N^2
        \left[
        \frac{2 H}{f^2}+\frac{P}{4 f^4}.
        \right]
    \label{EquationTimePRISM}
\end{eqnarray}
Note that for a STEM probe, the probe amplitude coefficients beyond the probe semi-angle are zero and so the number of beams $B$ used in practice is often much lower than the number of possible beams. The speedup offered by the PRISM algorithm is therefore approximately equal to the ratio of Eqs.~\ref{EquationTimeMulti} and \ref{EquationTimePRISM}, given by
\begin{equation}
    \label{EquationTimeRatio}
    \frac{T_{\rm{Multi}}}{T_{\rm{PRISM}}}
    = \frac{8 H P f^4}{B (8 H f^2+P)}.
\end{equation}
If the rate-limiting computation step for the PRISM algorithm is multiplying out the compact $S$-matrix, the speedup ratio does not depend on the number of probe positions $P$ and the speedup will vary with $f^4$. In the multislice and PRISM simulations given in the first results section below, the values of the terms of Eq.~\ref{EquationTimeRatio} were $H = 40$, $B \approx 10^4$ and $P \approx 10^5$. Plugging these numbers into Eq.~\ref{EquationTimeRatio} gives a speedup factor $T_{\rm{Multi}} / T_{\rm{PRISM}}$ of approximately 0.5, 8, 110 and 1100 for $f=$ 2, 4, 8 and 16 respectively.

\section*{Results and Discussion}

\subsection*{Comparison of Accuracy Between Multislice and PRISM}

\begin{figure*}[htbp]
    \centering
        \includegraphics[width=6.0in]{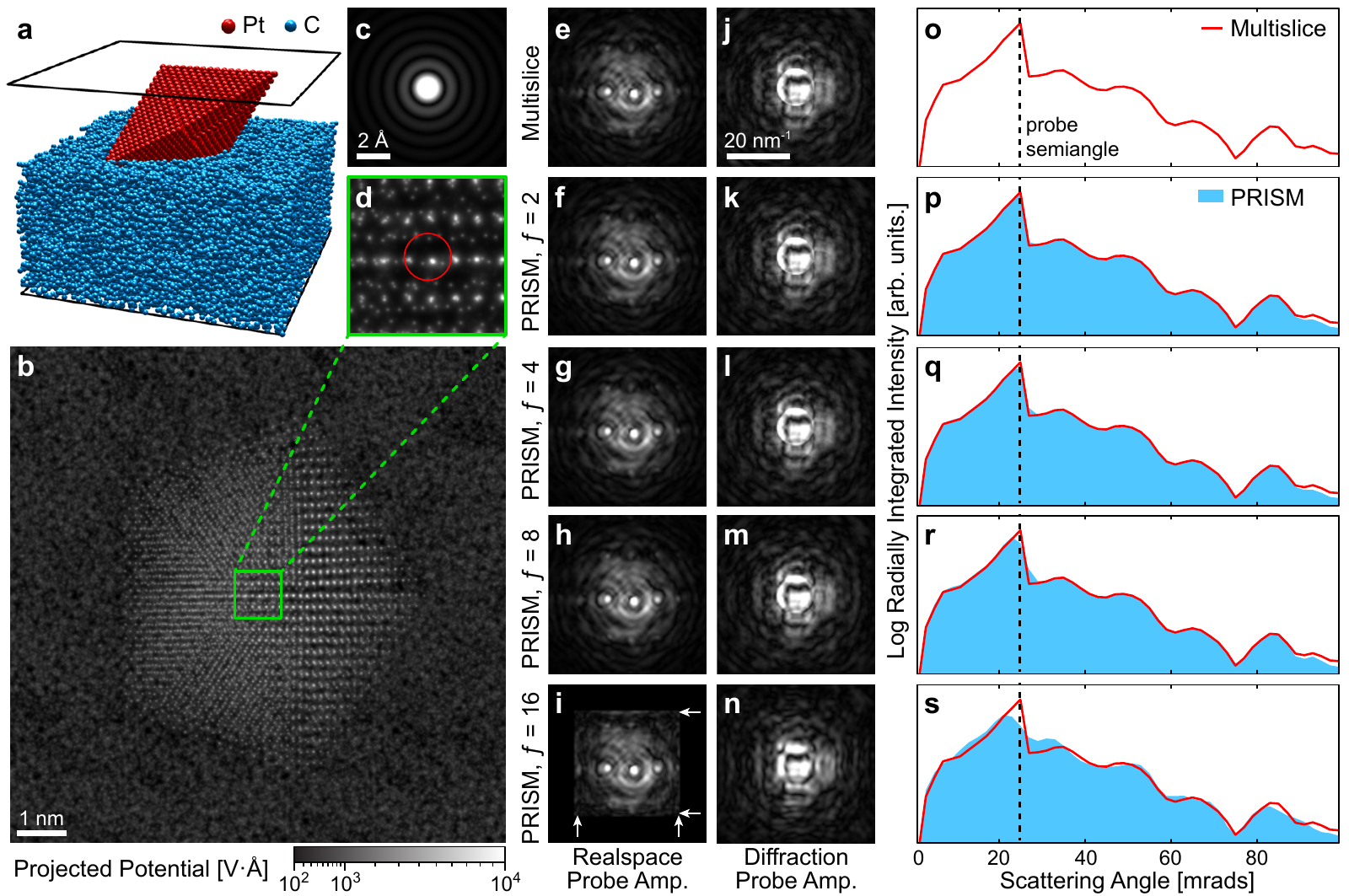}
 	\caption{Comparison of Multislice and PRISM simulations of a single converged electron probe. (a) Three-dimensional view of the atomic structure consisting of a defected Pt decahedral nanoparticle, resting up an amorphous carbon support. Entrance and exit planes shown in black. (b) Sum of projected potentials with inset around probe position shown in (d), and initial probe amplitude at the same location shown in (c). Realspace images of probe amplitude after passing through sample for (e) the multislice method, and (f)-(i) using various interpolation factors $f$ for the PRISM method. Diffraction space images of probe amplitude after passing through sample for (j) the multislice method, and (k)-(n) using various interpolation factors $f$ for the PRISM method. (o)-(s) Radially integrated intensities of (j)-(n) respectively, with multislice result overlaid in (p)-(s) for comparison.}
	\label{FigureCompMethods}
\end{figure*}

In general, PRISM will always be less accurate than corresponding multislice calculations, unless the PRISM speedup allows for finer pixel sampling, inclusion of higher scattering angles, or a similar improvement. However the increased error is negligibly small in many cases, and will depend heavily on the microscope and sample parameters of a given simulation. To demonstrate this, we have compared the accuracy of a STEM probe calculation for a typical experimental geometry: a Pt nanoparticle (NP) approximately 7 nm diameter tilted 30$\degree$ from the primary axis. The NP rests upon an amorphous carbon substrate with a thickness of 5 nm, shown in Fig.~\ref{FigureCompMethods}a.  The NP has a multiply-twinned decahedral structure, with screw and edge dislocations present in two of the grains. The NP atomic coordinates were taken from \cite{chen2013three}, and the amorphous carbon structure was adapted from \cite{ricolleau2013random}.

The sample was divided up into slices 0.2 nm thick, and the projected potential was computed for all slices. The sum of these potentials is shown in Fig.~\ref{FigureCompMethods}b, with an enlarged inset shown in Fig.~\ref{FigureCompMethods}d. The initial STEM probe generated from a 25 mrads semi-angle aperture at 80 kV is shown in Fig.~\ref{FigureCompMethods}c, with the probe center position shown in Fig.~\ref{FigureCompMethods}d. We then calculated the probe wavefunction after passing through the sample using the multislice method (Fig.~\ref{FigureCompMethods}e) and the PRISM algorithm with interpolation factors of $f=$ 2, 4, 8 and 16 (Figs.~\ref{FigureCompMethods}f-i). The corresponding probe amplitudes in Fourier space are shown in Fig.~\ref{FigureCompMethods}j-n respectively, and the logarithm of the radially integrated intensities are plotted in Figs.~\ref{FigureCompMethods}o-s respectively. In the real space images, the channeling effect along aligned atomic columns is visible in all simulations \cite{pennycook2011scanning}.

We see that the PRISM method correctly reproduces most of the fine structure in the real space probe images. In Fig.~\ref{FigureCompMethods}i, we see that when $f=16$ the tails of the probe have been cut off by the edge of the cropping window, leading to small artifacts at the boundary (shown with white arrows). However, Fig.~\ref{FigureCompMethods}n and Fig.~\ref{FigureCompMethods}s show that this simulation can still qualitatively reproduce the diffracted probe signal with good accuracy. 

Two small differences between the PRISM and multislice simulations are visible. The first is the ``blurring'' effect caused by the Fourier interpolation, an effect which increases as $f$ increases in Figs.~\ref{FigureCompMethods}k-n. This is reflected in the radially integrated intensities, as a small mixing between adjacent detector angle bins. Secondly, there is a small decrease in intensity at the highest scattering angles. This decrease is very small, visible only because of the logarithmic intensity scale. The source is probably the interpolation step of PRISM, which will reduce the image sharpness slightly, manifesting at the highest spatial frequencies / scattering angles. We therefore conclude that PRISM is accurate enough to replace the traditional multislice method for STEM simulations in most cases. The primary exceptions are when the probe is very large (highly defocused or delocalized) or when  fine details must be recovered from diffraction pattern, such as higher order Laue zone line measurements \cite{jones1977higher}.

\begin{figure*}[htbp!]
    \centering
        \includegraphics[width=6.0in]{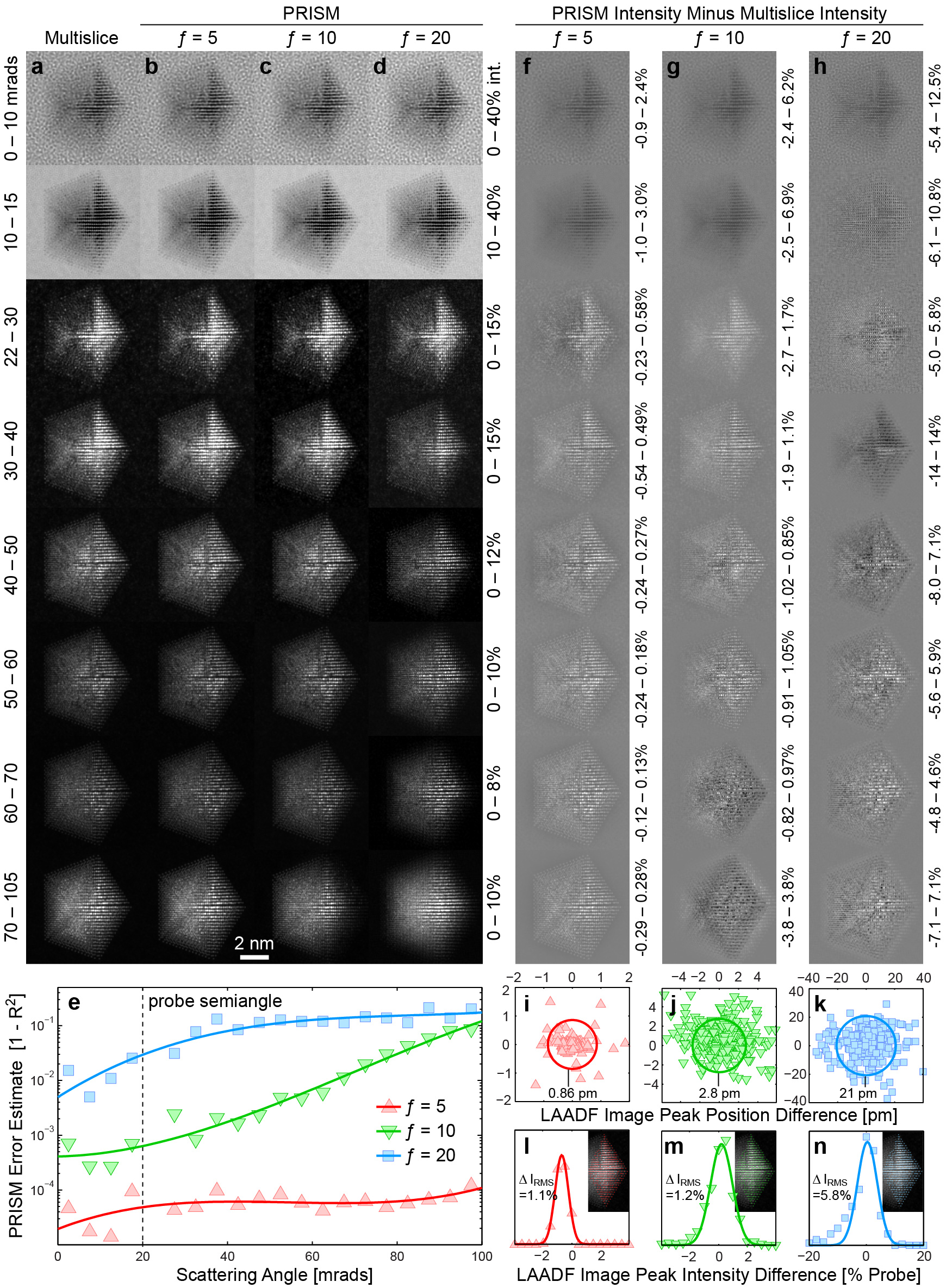}
 	\caption{STEM simulations of a Pt particle on amorphous carbon, using a 20 mrad STEM probe at 80 kV. (a) Multislice and (b)-(d) PRISM image simulations for interpolation factors of $f=5$, $10$, and $20$ respectively. Each row corresponds to a different annular virtual detector, with the inner and outer scattering angles labeled on the left. The intensity of each row was kept constant, in units of total probe intensity with the range shown to the right. (e) Error estimates as a function of scattering angle for the PRISM simulations in (b)-(d). (f)-(g) Intensity differences between PRISM and Multislice images, with plot ranges given to the right. (i)-(k) Peak positions differences and (l)-(m) peak intensity differences for 360 peaks fitted from LAADF images, between multislice simulations and PRISM simulations with $f=5$, $10$, and $20$ respectively. Mean position and RMS intensity differences, and the included peak positions are inset into (i)-(n).}
	\label{FigureCompImage}
\end{figure*}

To demonstrate the accuracy of PRISM, we have performed full image simulations of the sample shown in Fig.~\ref{FigureCompMethods}. Simulated STEM images are shown for various virtual detectors using the multislice method in Fig.~\ref{FigureCompImage}a. PRISM simulations using interpolation factors of $f=5$, 10 and 20 are plotted in Figs.~\ref{FigureCompImage}b-d respectively. The simulations correspond to cut out regions with side length 2, 1 and 0.5 nm respectively. Note that the annular detector inner angle in the third row of simulations in Fig.~\ref{FigureCompImage} is slightly increased to prevent sampling artifacts at the edge of the 20 mrad semiangle electron probe.

Fig.~\ref{FigureCompImage}a and b demonstrate that for relatively low interpolation factors, PRISM is essentially identical to Multislice simulations. PRISM can accurately capture the coherent diffraction image contrast present at lower scattering angles. Additionally it can reproduce the clean mass-thickness contrast signal present at high scattering angles. As the interpolation factor is increased, subtle differences from the multislice image simulations do emerge, in Figs~\ref{FigureCompImage}c-d. However, the image contrast is still qualitatively very similar to the multislice images. The primary advantage of PRISM is the reduced calculation time; the PRISM simulations with interpolation factors of $f=5$, 10 and 20 gave speed up factors of approximately 40, 280, and 2100 respectively compared to the multislice simulation. The $f=20$ simulation shown in Fig.~\ref{FigureCompImage}d requires only a few minutes of calculation time on a modern desktop computer, using Matlab code that has not been highly optimized or compiled.

Fig.~\ref{FigureCompImage}e shows an error estimate for the three PRISM simulations in Fig.~\ref{FigureCompImage}b-d.  The error was estimated as $1-R^2$, where $R^2$ is the correlation coefficient between the multislice and PRISM simulation pixel intensities. The $f=5$ PRISM simulation error is approximately 0.005\% for all scattering angles, indicating that this simulation is essentially error-free. When the interpolation factor is increased to $f=10$, the difference from a multislice simulation increases to an error of 0.05\% for low scattering angles and $\approx 1\%$ error for intermediate scattering angles, and finally $\approx$10\% for high scattering angles. Doubling the interpolation factor again to $f=20$ gives error of 1\% error at small scattering angles and 10\% error for medium and high scattering angles. This larger error is caused by the region cropped around the STEM probe being small enough to crop out a significant portion of the probe intensity and cause boundary errors, as in Fig.~\ref{FigureCompMethods}i. We conclude that when using a low enough interpolation factor $f$, the PRISM method can simulate STEM intensities at all scattering angles with negligibly low error. The best value for $f$ can be determined by testing probes at different locations in the simulation cell with both PRISM and multislice, or by using a conservative, low value; for example in this simulation $f=5$ leads to a cutout region with side length 2 nm, large enough to contain the entire STEM probe for any probe semi-angle large enough to generate atomic resolution contrast.

Figs.~\ref{FigureCompImage}f-h show the difference in intensities between the PRISM image simulations in Figs.~\ref{FigureCompImage}b-d respectively, and the multislice image simulations in Figs.~\ref{FigureCompImage}a. The intensity range for each panel is set individually to show good contrast for the features present. Fig.~\ref{FigureCompImage}f shows that when $f$ is small, PRISM will slightly over-estimate the image intensity at scattering angles below the probe semiangle, and slightly under-estimates the intensity at higher scattering angles. These intensity differences are probably caused by the different sampling of PRISM compared to multislice for both defining the initial electron probe, and creating the virtual detectors for the output signal. The errors for $f=5$ also appear to be primarily intensity errors, which will not strongly affect measurements such as peak position estimation. Figs.~\ref{FigureCompImage}g and h show larger intensity differences for $f=10$ and $20$. These differences depend on the amount of local scattering and the local tilts of atomic columns, which could introduce errors in peak position measurements.

To test the accuracy of using PRISM to estimate structural information, we have used non-linear least squares peak fits using a 2D Gaussian function to measure 360 of the strongest peaks on the right-hand side of the decahedral particle in Fig.~\ref{FigureCompImage}. These peaks were measured from low angle annular dark field (LAADF) images created with a virtual detector from 22.5 to 105 mrads. We have plotted the differences in measured peak positions between multislice and PRISM simulations in Figs.~\ref{FigureCompImage}i-k, and the peak intensity differences in Figs.~\ref{FigureCompImage}l-m. The mean 2D position errors for the PRISM simulations are 0.86, 2.8 and 21 pm for $f=5$, $10$ and $20$ respectively. These errors will decrease if more frozen phonon configurations are included due to the increasing smoothness of the peak functions in the simulated images. Additionally the errors could be reduced by using a probe sampling finer than $0.25 \rm{\AA}$. The intensity differences in the peaks are fairly small, $\approx 1\%$ for both $f=5$ and $10$. When $f$ is increased to 20 the intensity errors increase rapidly, underlining the importance of choosing an $f$ value low enough for the desired accuracy.

\subsection*{PRISM Simulations with Varying Probe Size}

In the PRISM method, once the compact $S$-matrix is computed for a given set of atomic coordinates, it can be used for many different simulations. The primary change between electron probes is the probe center position, but we can also vary coherent wave aberrations in the probe such as defocus or spherical aberration, change the probe size by modifying the probe semi-angle radius, and also include relative tilt between the probe and sample by moving the probe center away from $\vec{q}=(0,0)$. These simulation parameter changes reflect only changes in the probe weighting coefficients $\alpha_{m,n}(\vec{r}_0)$, given in Eq.\ref{EquationPRISMfinal}.

\begin{figure}[htbp!]
    \centering
        \includegraphics[width=3.2in]{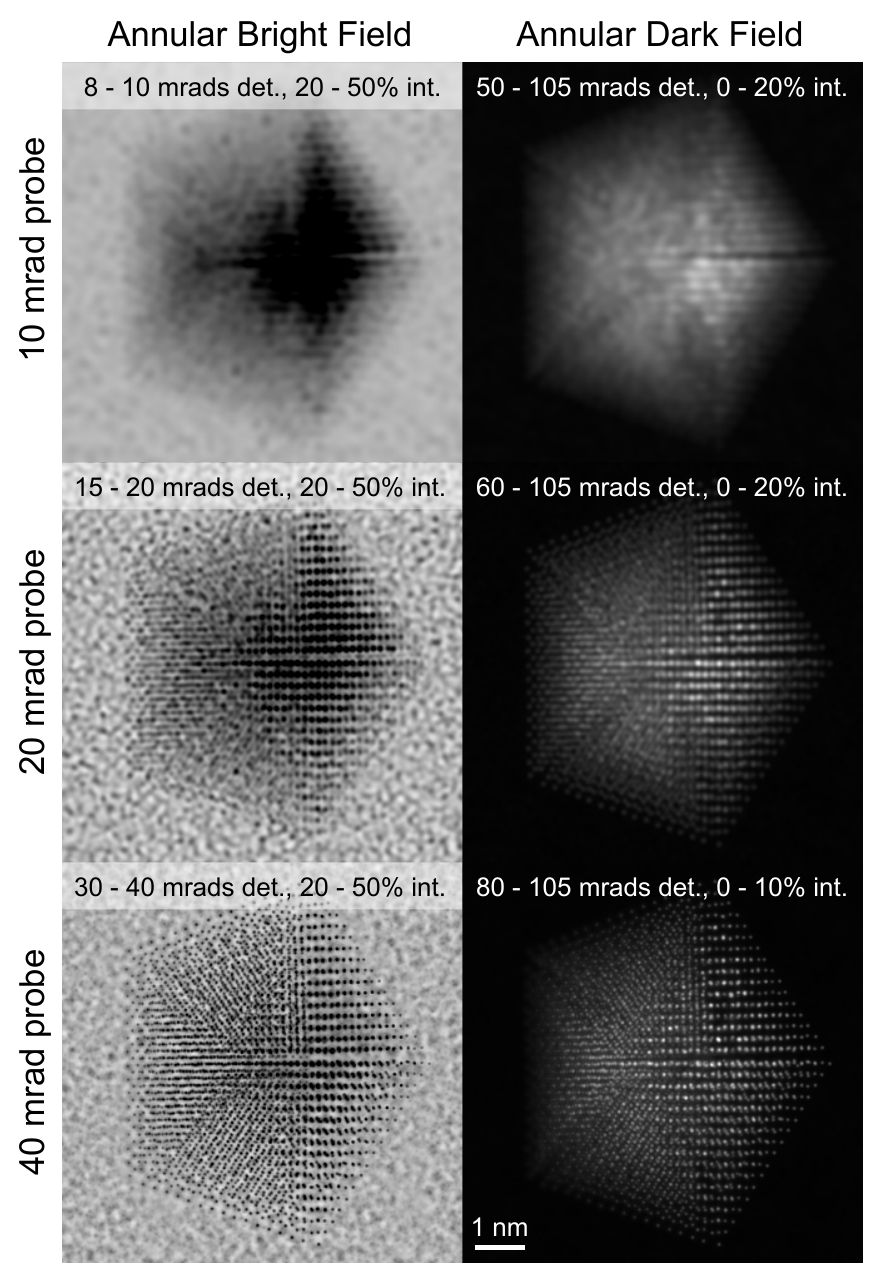}
 	\caption{STEM image simulations of a Pt decahedral nanoparticle sitting on an amorphous carbon substrate. Images simulated with the PRISM algorithm for probe semi-angles of 10, 20 and 40 mrads, for annular detectors that generate a bright field and dark field contrast.}
	\label{FigureProbeSize}
\end{figure}

An example of using the same $S$-matrix to simulate STEM images with varying probe size and annular detectors is shown in Fig.~\ref{FigureProbeSize}, for an accelerating voltage of 80 kV and a probe spacing of 0.025 nm. Based on the previous results shown in Fig.~\ref{FigureCompImage}, we have chosen an interpolation factor of $f=5$ for these simulations. In Fig.~\ref{FigureProbeSize} we have generated annular bright field images by setting the detector inner and outer angles to $\approx$75\% and 100\% of the probe semi-angle respectively. Annular dark field images were generated by setting the detector inner angle to 40 mrads outside of the probe semi-angle. These simulations show that for this sample, a 10 mrad probe does not generate atomic resolution contrast. However, a 20 mrad probe can resolve the atomic columns on the two grains on the right hand side of Fig.~\ref{FigureProbeSize}. Resolving atomic columns over the entire particle requires increasing the probe semi-angle to 40 mrads.

\newpage
\section*{Conclusion}

In summary, we have presented the PRISM algorithm for STEM image simulation, which combines aspects of the Bloch wave and multislice simulation methods. PRISM uses Fourier interpolation with an integer factor $f$, and can lead to a a decrease in computation time that is proportional to $f^4$ in many cases. We have compared PRISM and multislice image simulations and shown that as long as $f$ is kept small enough, the simulation error for PRISM is negligibly small. Large $f$ values can be used to generate a rough contrast model for a given simulation cell in very short computation times.  We expect that the PRISM method will find wide application in STEM studies that require image simulation, due to its potential for a large speed up relative to the multislice method.


\begin{backmatter}

\section*{Declarations}

\section*{Authors' contributions}

CO conceived of the PRISM method, implemented the multislice and  PRISM methods, simulated and analyzed the results, and wrote the manuscript.

\section*{Acknowledgements}

We thank Earl Kirkland, Christoph Koch and Roar Kilaas for helpful discussions about (S)TEM simulation methods. We also thank Hao Yang, Jim Ciston, Tyler Harvey and Peter Ercius for helpful suggestions on this manuscript.

\section*{Funding}

 Work at the Molecular Foundry was supported by the Office of Science, Office of Basic Energy Sciences, of the U.S. Department of Energy under Contract No. DE-AC02-05CH11231.

\section*{Competing interests}

The authors declare that they have no competing interests.

\section*{Availability of data and material}

Please contact the author for code examples and updates.

\section*{Ethics approval and consent to participate}

Not applicable.

\section*{Consent for publication}

I consent for this manuscript to be published under the Creative Commons Attribution 4.0 International License.

\bibliographystyle{bmc-mathphys} 
\bibliography{PRISMrefs}


\end{backmatter}

\end{document}